\documentclass[a4paper]{article}
\usepackage{INTERSPEECH2022}
\usepackage{multirow}
\usepackage{makecell}
\usepackage{url}

\title{A Multi-grained based Attention Network for Semi-supervised Sound Event Detection}
\name{Ying Hu$^1$$^2$, Xiujuan Zhu$^1$$^2$,Yunlong Li$^1$$^2$, Hao Huang$^1$$^2$, and Liang He$^2$$^3$}
\address{
	$^1$Key Laboratory of signal detection and processing in Xinjiang, China\\
	$^2$School of Information Science and Engineering, Xinjiang University, Urumqi, China\\
$^3$Department of Electronic Engineering, Tsinghua University, China}
\email{huying@xju.edu.cn, xjzhu@stu.xju.edu.cn, liyunlong@stu.xju.edu.cn}

\begin{document}

\maketitle
\begin{abstract}
Sound event detection (SED) is an interesting but challenging task due to the scarcity of data and diverse sound events in real life. 
This paper presents a multi-grained based attention network (MGA-Net) for semi-supervised sound event detection. 
To obtain the feature representations related to sound events, a residual hybrid convolution (RH-Conv) block is designed to boost the vanilla convolution's ability to extract the time-frequency features. 
Moreover, a multi-grained attention (MGA) module is designed to learn temporal resolution features from coarse-level to fine-level. With the MGA module, the network could capture the characteristics of target events with short- or long-duration, resulting in more accurately determining the onset and offset of sound events. 
Furthermore, to effectively boost the performance of the Mean Teacher (MT) method, a spatial shift (SS) module as a data perturbation mechanism is introduced to increase the diversity of data. Experimental results show that the MGA-Net outperforms the published state-of-the-art competitors, achieving 53.27$\%$ and 56.96 $\%$ event-based macro F1 (EB-F1) score, 0.709 and 0.739 polyphonic sound detection score (PSDS) on the validation and public set respectively. 
\end{abstract}
\noindent\textbf{Index Terms}: Sound Event Detection, Semi-supervised Learning, Multi-grained Attention

\section{Introduction}
Sound event detection (SED) aims to detect the onset and offset of sound events and identify the class of target events.
Recently, there has been an increasing interest in semi-supervised SED in the Detection and Classification of Acoustic Scenes and Events (DCASE) challenge Task4 \footnotemark[1]. Sound event detection has wide applications, including audio surveillance systems \cite{valenzise2007scream}, monitoring systems \cite{debes2016monitoring} and smart homes\cite{turpault2019sound}.
\footnotetext[1]{\url{https://dcase.community/challenge2019/task-sound-event-detection-in-domestic-environments}.}

In the real world, different sound events exhibit unique patterns reflected in the time-frequency distribution. As a consequence,  
it is necessary to obtain the effective feature representation related to sound events. 
Thanks to the development of deep learning approaches, recent advances \cite{parascandolo2016recurrent, cakir2017convolutional} have led to improved performance in SED task. 
Several standard convolutional neural network (CNN) blocks were stacked as the feature encoder to generate the high-level feature representations for the SED task \cite{lin2020specialized, huang2020multi}. Lu et al. \cite{lu2018multi} proposed a multi-scale recurrent neural network (RNN) to capture the fine-grained and long-term dependencies of sound events. CNN is good at learning features shifted in both time and frequency, while RNN models longer temporal context information.

Convolutional recurrent neural network (CRNN) approaches have shown their superiority in the estimation of onset and offset \cite{yan2020task, dinkel2021towards}.
For better-integrating information from different time resolutions, Guo et al. \cite{guo2019multi} proposed multi-scale CRNN to learn coarse or fine-grained temporal features by applying multiple RNNs.
Recently, some works \cite{miyazaki2020weakly, Miyazaki2020} also proposed to combine CNN with the self-attention mechanism for the SED task that instead of applying RNN, that self-attention mechanism is used to model temporal context information. To be specific, Miyazaki et al. \cite{miyazaki2020weakly} incorporated the self-attention mechanism of the Transformer in SED to capture global time features and had shown its superior performance in SED. Then they further proposed the Conformer-based SED method \cite{Miyazaki2020} to capture both global and local time context information of an audio feature sequence simultaneously.

\begin{figure}[t]
	\centering
	\includegraphics[scale=.38]{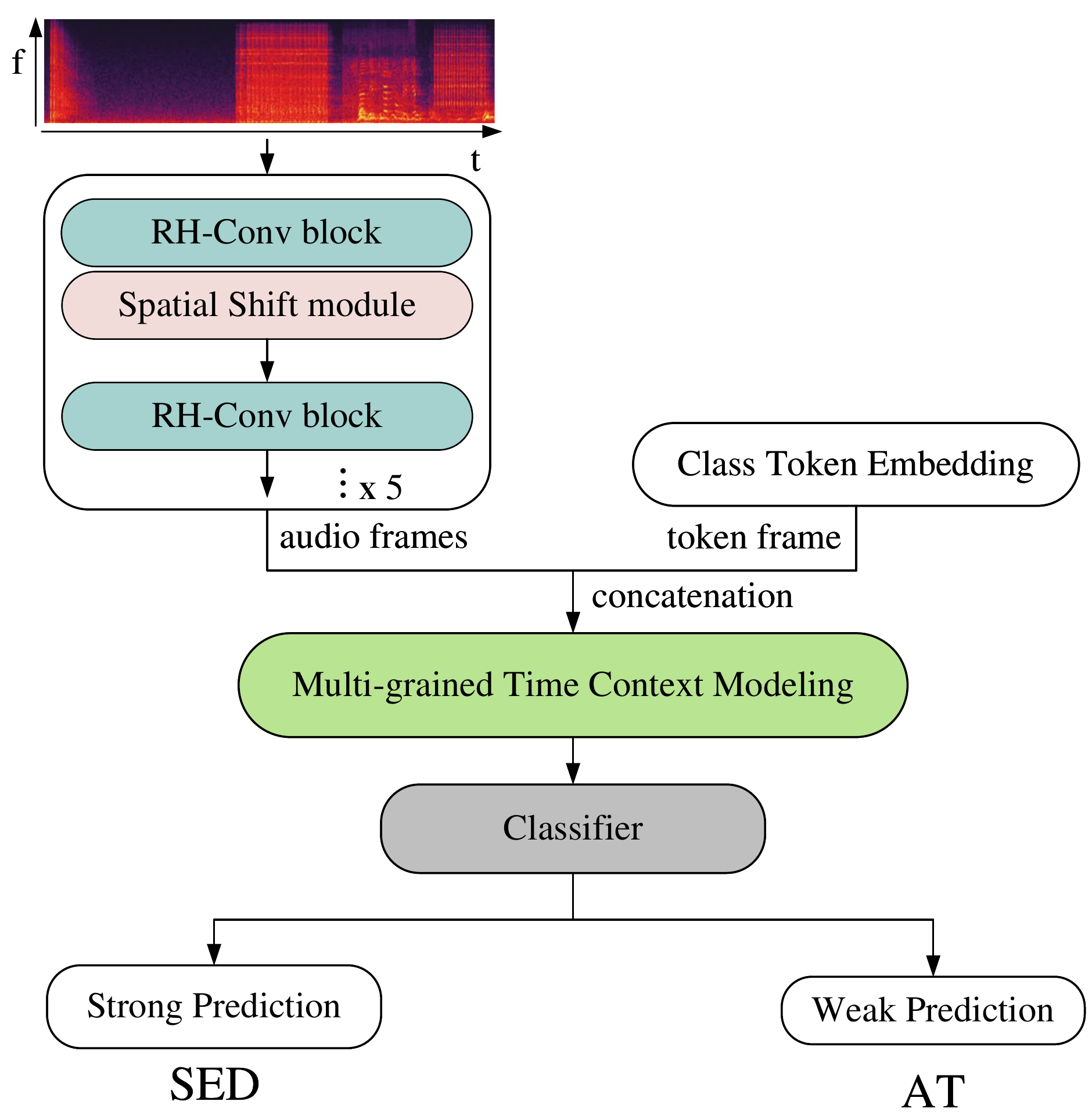}
	\caption{Illustration of the proposed MGA-Net.}
	\label{fig1}
\end{figure}

In addition, similar to \cite{jiakai2018mean, turpault2020training}, Mean Teacher \cite{tarvainen2017mean} method is adopted to perform semi-supervised learning (SSL) for SED in this paper. 
Under the cluster assumption that two samples close to each other in the input feature space are likely to belong to the same class \cite{luo2018smooth}, some SSL methods \cite{laine2016temporal, tarvainen2017mean, yan2020task} introduced a consistency regularization based on perturbation techniques. 
Data perturbation methods \cite{verma2019interpolation, miyato2018virtual} play an essential role in introducing effective perturbation for SSL learning.
Zheng et al. \cite{zheng2020effective} also showed that the MT method could benefit from suitable data and/or model perturbation.

\begin{figure}[t]
	\centering
	\includegraphics[scale=.35]{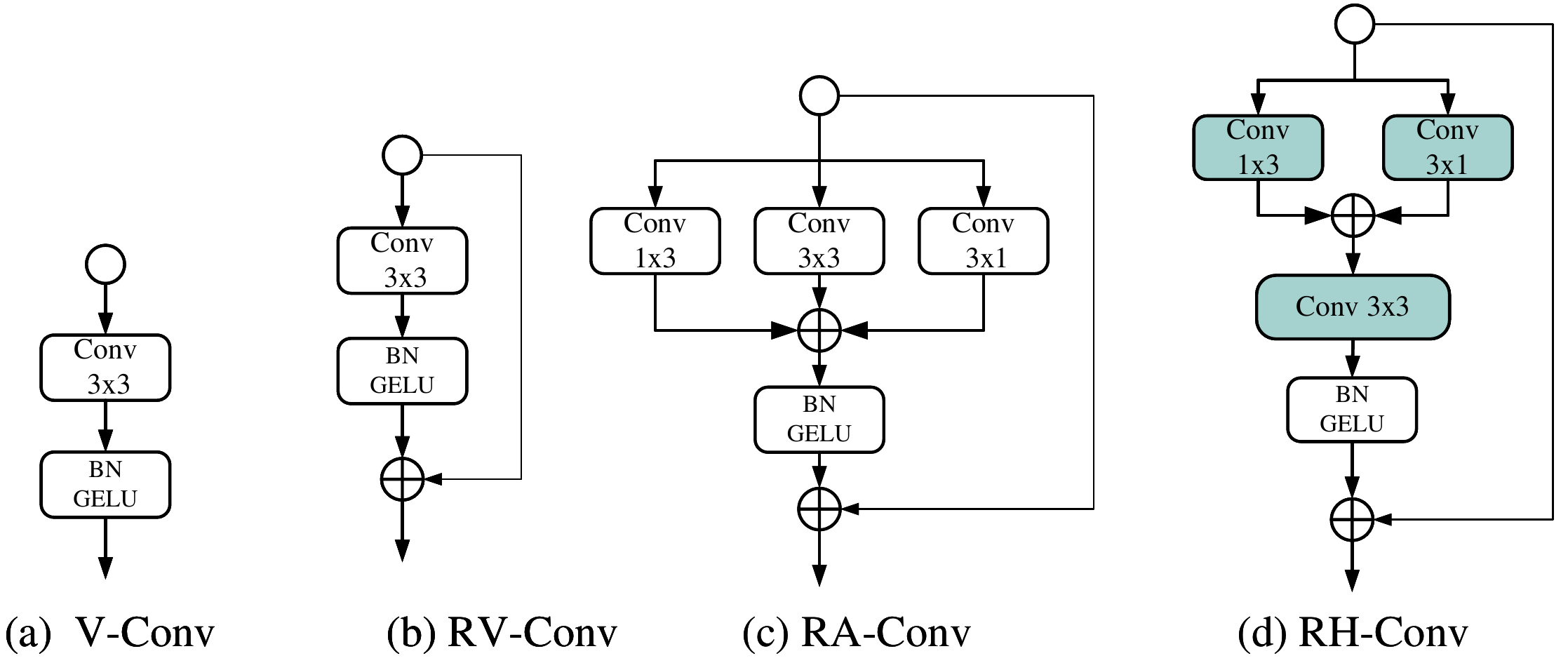}
	\caption{Four kinds of CNN feature extraction blocks. (a) Vanilla convolution block (V-Conv). ( b) Residual Vanilla convolution (RV-Conv) block. (c) Residual Asymmetric convolution (RA-Conv) block. (d) Residual Hybrid convolution (RH-Conv) block.}
	\label{fig2}
\end{figure}

Inspired by the above-mentioned works, we propose a multi-grained based attention network (MGA-Net) in this paper.
For the time-frequency feature extraction, we explore four kinds of feature extraction blocks based on CNN and design residual hybrid convolution (RH-Conv) block to boost the representation power of vanilla convolution.
We also design a multi-grained based attention (MGA) module to utilize the temporal information. The MGA module builds upon three stages of feature learning: global, local, and frame-level time context modeling. 
It can capture well the features of temporal resolution from coarse to fine-level.
Similar to data augmentation, which can increase the diversity of data, a spatial-shift module is designed as a data perturbation mechanism to bring about data augmentation for the MT method. 
Experiments on the dataset of DCASE 2020 task4 demonstrate the superiority of our proposed methods.

\begin{figure}[b]
	\centering
	\includegraphics[width=\linewidth]{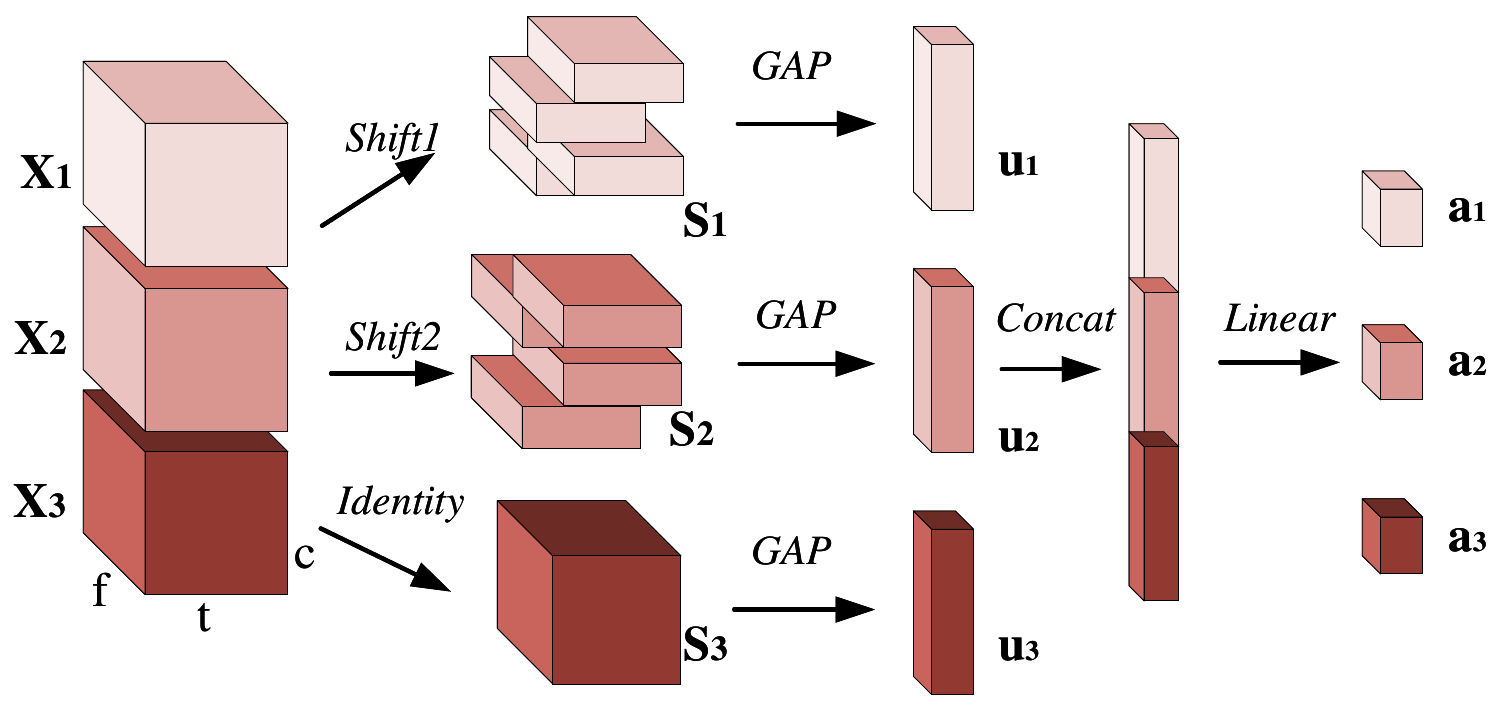}
	\caption{Illustration of the proposed Spatial Shift module.}
	\label{fig3}
\end{figure}

\section{Proposed Method}

Our proposed MGA-Net is shown in Fig.~\ref{fig1}. 
It employs six residual hybrid convolution blocks and one spatial shift module to extract time-frequency features, where each residual hybrid block is followed by an average pooling and dropout layer. Then the extracted features are fed into the multi-grained time context modeling to learn the temporal context information. A linear classifier based on a dense layer with sigmoid activation is followed to perform strong label prediction for the SED detection task. Similar to \cite{miyazaki2020weakly}, a class token embedding is used to aggregate the whole sequence information that performs weak label prediction for the audio tagging (AT) classification task.
The following subsections will describe the RH-Conv block, SS module, and MGA module.

\subsection{Residual Hybrid Convolution Block}
We build four kinds of CNN feature extraction blocks as shown in Fig.~\ref{fig2}. 
Each CNN layer is followed by batch normalization (BN) and gaussian error linear unit (GELU) \cite{hendrycks2016gaussian} activation.
Fig.~\ref{fig2} (a) is the vanilla CNN with square kernels, i.e.,  3$\times$3, referred to as “V-Conv”. Fig.~\ref{fig2} (b), referred to as “RV-Conv”, introduces identity mapping as the residual connection based on the “V-Conv” block. Fig.~\ref{fig2} (c) can be viewed as asymmetric convolution \cite{ding2019acnet} comprising three parallel CNN layers with 3$\times$3, 1$\times$3 and 3$\times$1 kernels, respectively, referred to as “RA-Conv”.
Fig.~\ref{fig2} (d) is our proposed residual hybrid convolution block, which is a combination of using two parallel CNN layers with 1$\times$3 and 3$\times$1 kernels followed by vanilla convolution with 3$\times$3 kernels. It applies two asymmetric convolution kernels to strengthen the square convolution kernels and is referred to as the “RH-Conv” block. Four kinds of feature extraction blocks are explored with the goal of designing a better CNN structure to extract more robust features related to sound events.

\subsection{Spatial Shift Module}

To provide a data perturbation mechanism for the MT semi-supervised method,  we design a spatial shift module. It firstly conducts the spatial-shift operation, which is proposed by \cite{Yu2021S2MLPv2IS}, helping to increase the diversity of features. And it further evaluates the degree of importance for spatial shift operation by generating the corresponding weights.

Given an input feature map $\textbf{X}$ $\in R^{C\times T\times F}$, we firstly expand the channels of $\textbf{X}$ from $c$ to 3$c$ by a linear layer. Then the expanded feature map is equally splitted into three parts: $\textbf{X}_{i}$ $\in R^{C\times T\times F}$ $i$=1, 2, 3. As shown in Fig.~\ref{fig3},  $\textbf{X}_{1}$ and $\textbf{X}_{2}$ are shifted as $\textbf{S}_{1}$ and $\textbf{S}_{2}$ through the $Shift1$ and $Shift2$ operation, respectively. $Shift1$ conducts the shift operations along the time and frequency dimension, respectively, as shown in Equation~\ref{eq1}. In contrast, $Shift2$ conducts an asymmetric spatial-shift operation with respect to $Shift1$ as shown in Equation~\ref{eq2}. Thus, they are complementary to each other.  $\textbf{X}_{3}$ is just identified as $\textbf{S}_{3}$. Then, we embed the global information vector by using global average pooling on $\textbf{S}_{i}$. The global vectors $\textbf{u}_{i}$ $\in R^{C\times 1\times 1}$ are concatenated together along the channel dimension. A linear layer is followed to generate weights $\textbf{a}_{i}$, which is used to reweigh $\textbf{S}_{i}$. Then the softmax function is applied on the weights $\textbf{a}_{i}$ to limit $\sum_{i=1}^{i=3}\textbf{a}_i=1$. In all, the final output $\ \textbf{X}_{out}$ $\in R^{C\times T\times F}$ of this module can be writing as 
$\ \textbf{X}_{out}=\sum_{i=1}^{i=3}{\textbf{a}_i \times  \textbf{S}_i}$.

\begin{equation}
	\begin{aligned}
		\textbf{X}_1[1:t,:,:c/4]&\gets \textbf{X}_1[0:t-1,:,:c/4];\\
		\textbf{X}_1[0:t-1,:,c/4:c/2]&\gets \textbf{X}_1[1:t,:,c/4:c/2];\\
		\textbf{X}_1[:,1:f,c/2:3c/4]&\gets \textbf{X}_1[:,0:f-1,c/2:3c/4];\\
		\textbf{X}_1[:,0:f-1,3c/4:]&\gets \textbf{X}_1[:,1:f,3c/4:]
	\end{aligned}
	\label{eq1}
\end{equation}

\begin{equation}
	\begin{aligned}
		\textbf{X}_2[:,1:f,:c/4]&\gets \textbf{X}_2[:,\ 0:f-1,:c/4];\\
		\textbf{X}_2[:,0:f-1,c/4:c/2]&\gets \textbf{X}_2[:,\ 1:f,c/4:c/2];\\
		\textbf{X}_2[1:t,:,c/2:3c/4]&\gets \textbf{X}_2[0:t-1,:,c/2:3c/4];\\
		\textbf{X}_2[0:t-1,:,,3c/4:]&\gets \textbf{X}_2[1:t,:,3c/4:]
	\end{aligned}
	\label{eq2}
\end{equation}

\begin{figure}[t]
	\centering
	\includegraphics[width=\linewidth]{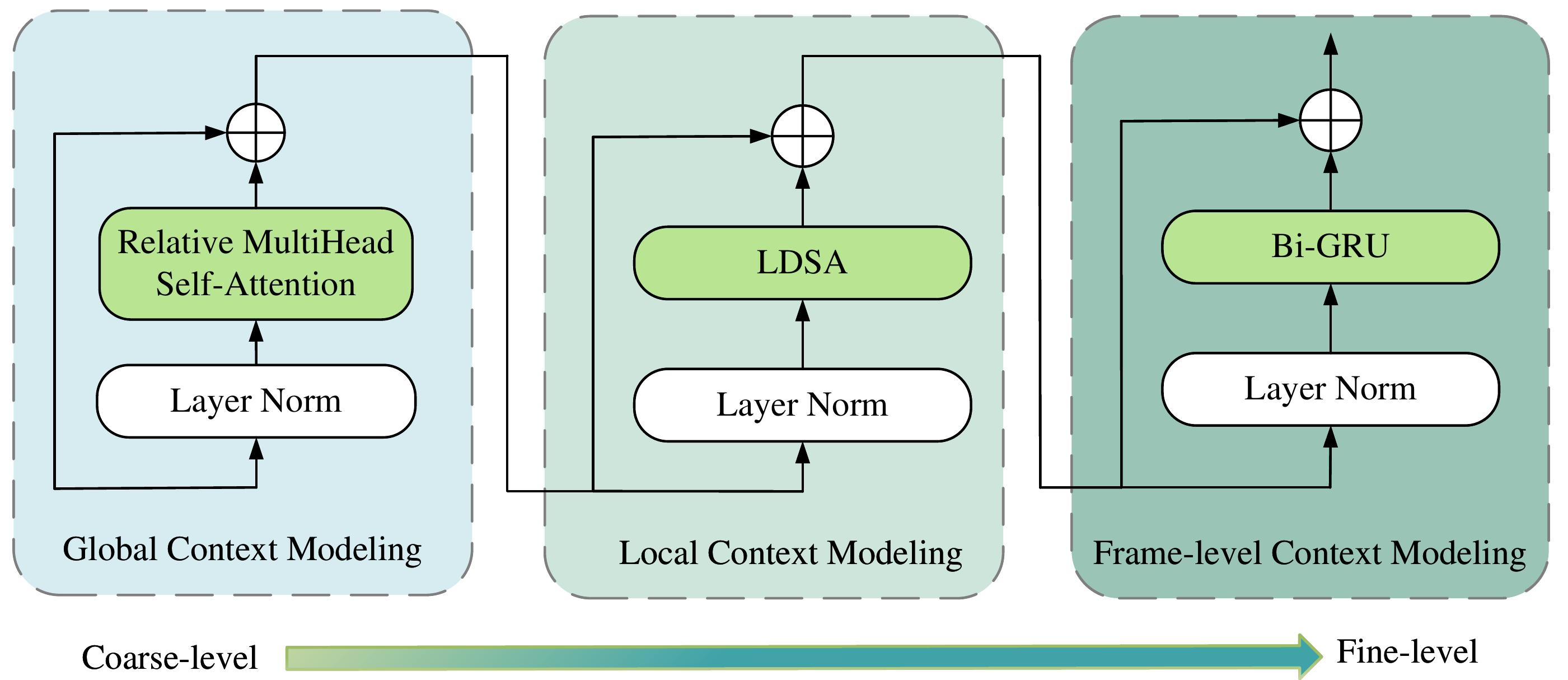}
	\caption{Illustration of the proposed Multi-grained Attention module. The green arrow denotes the time context is modeled from coarse to fine-level, and conversely, is modeled from fine to coarse-level.}
	\label{fig4}
\end{figure}

\subsection{Multi-Grained Attention Module}

The multi-grained based attention module is designed to model the temporal context dependencies from coarse-level to fine-level. As shown in Fig.~\ref{fig4}, there are three main processes in the multi-grained attention module: Global Context Modeling, Local Context Modeling, Frame-level Context Modeling.
We also add residual connection and layer normalization ($LN$) operation at each modeling process.

\subsubsection{Global Context Modeling}

The global context modeling is built upon the multi-head self-attention mechanism \cite{vaswani2017attention}. Considering the sequential position of input features, we introduce relative positional encoding (RPE) \cite{dai2019transformer} which has been shown effective in SED task \cite{Sundar2021EventSA} to encode position information of inter-frames.
The length of attention weights is that of the entire time series, making the feature representation more global but coarser. Assuming the input sequence is \textbf{X} $\in R^{T\times d}$, the global context modeling can be written as:

\begin{equation}
	\textbf{X}_{global} = RA(LN(\textbf{X}))+\textbf{X}
\end{equation}

Where $RA$ denotes the multi-head self-attention with relative positional encoding and $LN$ the layer normalization.
\subsubsection{Local Context Modeling}
Local context modeling is designed to capture the local time
dependencies within specific time frames rather than the entire time series, complementing the global context modeling. We use local dense synthesizer attention (LDSA) \cite{xu2021transformer} to achieve local context modeling. The local context modeling is expressed as follows:

\begin{equation}
	\textbf{X}_{local} = LDSA(LN(\textbf{X}_{global} ))+\textbf{X}_{global}
\end{equation}
The LDSA firstly defines a context window $c$ which
restricts the attention scope to a local range around the current central frame. Attention weights of the other frames outside the context width are set to 0. $c$ is set to 3 in our experiment.
The current frame is restricted to only interact with its finite neighbouring frames, thus, achieving the learning of local features. The process of LDSA is calculated as follows:
\begin{equation}
	A(\textbf{X}_{global}) = Softmax(\sigma (\textbf{X}_{global}\textbf{W}_{1})\textbf{W}_{2})
\end{equation}

\begin{equation}
	\textbf{V}=\textbf{X}_{global}\textbf{W}_{3}
\end{equation}
where $\textbf{W}_{1}$ $\in R^{d\times d}$, $\textbf{W}_{2}$ $\in R^{d\times c}$ and $\textbf{W}_{3}$ $\in R^{d\times d}$ are learnable weights.

Then it assigns the attention weights to the current frame and its neighboring frames:
\begin{equation}
	\textbf{Y}_{t}={\textstyle \sum_{j=0}^{c-1}}A_{(t,j)}(\textbf{X}_{global})\textbf{V}_{t+j-\lfloor c/2\rfloor} 
\end{equation}
Thus, the finally output of LDSA is obtained by:
\begin{equation}
	LDSA(X)=[\textbf{Y}_{0}, ..., \textbf{Y}_{t}, ..., \textbf{Y}_{T}]\textbf{W}^{o} 
\end{equation}
where the $\textbf{W}^{o}$ $\in R^{d\times d}$ is learnable weight.

\subsubsection{Frame-level Context Modeling}

No matter the global or local context modeling, the close correlation among time frames is lacking. Thus, we introduce frame-level context modeling to learn the fine-grained inter-frame features. 
Compared to the self-attention mechanism, RNN can directly model the sequential information naturally present in a sequence of frames. We use Bi-GRU to perform frame-by-frame detection and capture the long-term context dependencies for both past and future frames of the time series.
The calculation process is as follows:
\begin{equation}
\textbf{X}_{frame} = Linear(\sigma(BiGRU(LN(\textbf{X}_{local}))+\textbf{X}_{local})
\end{equation}
Where the $\sigma$ denotes ReLU activation function.

\section{Experiment Setup}

\subsection{Dataset}

The experiments in this paper were conducted on the dataset of task 4 in the DCASE2020. It has ten classes of sound events from the domestic environment. The dataset contains three types of training data: weakly labeled data (1502 clips), unlabeled data (13723 clips), and strongly labeled data (2584 clips). 
We evaluate the performance of the SED network on the validation (1083 clips) and public (692 clips) set.

The input features were Log-Mel spectrograms extracted from the 10-sec audio clips resampled to 16000 Hz.
The Log-Mel spectrogram was computed over 1024-point STFT windows with a hop size of 323 samples and 64 Mel-scale filters, resulting in an input feature matrix with 496 frames and 64 Mel-scale filters. More details of preprocessing and post-processing schemes used in our experiments were consistent with that setting in \cite{Miyazaki2020}.

\subsection{Experimental settings}

Our proposed MGA-Net was trained using the RAdam optimizer \cite{liu2019variance}, where the initial learning rate was set to 0.001. The size of the average pooling layer is set to 2$\times$2 in the first two layers and 1$\times$2 in the rest layers. The dropout rate was 0.1. In the multi-grained time context modeling, we applied 4 multi-grained attention modules, in which the dimension of features $d$ was set to 144, the number of attention heads 4, and the hidden size of the Bi-GRU 512. 
The loss function is a weighted sum of the classification and consistency losses. The classification loss based on binary cross-entropy (BCE) is calculated by the predictions and the ground truth, while the consistency loss is based on the mean squared error (MSE) between the outputs of student and teacher network.
Event-based macro F1 (EB-F1) \cite{mesaros2016metrics} and polyphonic sound detection score (PSDS) \cite{bilen2020framework} are used as the main evaluation metrics.

\begin{table}[th] \centering
	\caption{Performance comparison between the proposed MGA-Net and  the state-of-the-art SED methods. SS denotes the spatial shift module. }
	\renewcommand\arraystretch{1.2}
	\renewcommand\tabcolsep{4.5pt}
	\label{table1}
	\begin{tabular}{l|ll|ll}
		\Xhline{0.8pt}
		\multicolumn{1}{c|}{\multirow{2}{*}{\textbf{Method}}} & \multicolumn{2}{c|}{\textbf{Validation}} & \multicolumn{2}{c}{\textbf{Public}} \\ \cline{2-5} 
		\multicolumn{1}{c|}{}                                 & \textbf{EB-F1}      & \textbf{PSDS}      & \textbf{EB-F1}   & \textbf{PSDS}    \\ \hline
		Conformer-SED {\cite{Miyazaki2020}}                                  & 47.70               & 0.637              & 49.00            & 0.681            \\
		ESA-Net {\cite{Sundar2021EventSA}}                      & 47.80               & 0.688              & 52.10            & 0.712            \\ \hline
		\midrule
		MGA-Net(Coarse-Fine)                                              & \textbf{53.27}      & \textbf{0.709}     & \textbf{56.96}   & \textbf{0.739}   \\
		\quad -SS                                                & 52.43               & 0.705              & 56.48            & 0.737            \\ \Xhline{0.8pt}
	\end{tabular}
\end{table}

\begin{table}[bh] \centering
	\caption{Comparison among four kinds of feature extraction blocks.}
	\renewcommand\arraystretch{1.2}
	\renewcommand\tabcolsep{7pt}
	\label{table2}
	\begin{tabular}{l|ll|ll}
		\Xhline{0.8pt}
		\multicolumn{1}{c|}{\multirow{2}{*}{\textbf{Method}}} & \multicolumn{2}{c|}{\textbf{Validation}} & \multicolumn{2}{c}{\textbf{Public}} \\ \cline{2-5} 
		\multicolumn{1}{c|}{}                                 & \textbf{EB-F1}      & \textbf{PSDS}      & \textbf{EB-F1}   & \textbf{PSDS}    \\ \hline
		V-Conv                                                & 51.22               & 0.690              & 54.88            & 0.728            \\
		RV-Conv                                               & 52.31               & 0.703              & 55.31            & 0.728            \\
		RA-Conv                                               & 52.08               & 0.698              & 56.18            & 0.726            \\
		RH-Conv                                               & \textbf{52.43}      & \textbf{0.705}     & \textbf{56.48}   & \textbf{0.737}   \\ \Xhline{0.8pt}
	\end{tabular}
\end{table}

\section{Results and Discussion}
To investigate the effectiveness of the proposed MGA-Net, we compare it with the state-of-the-art methods \cite{Miyazaki2020, Sundar2021EventSA}. As shown in Table~\ref{table1}, the MGA-Net achieves 53.27$\%$, and 56.96$\%$ EB-F1 score, 0.709 and 0.739 PSDS score for the validation and public set, respectively, significantly outperforming the compared methods. 
In addition, by removing the spatial shift (SS) module, the network performance degrades slightly on both datasets. 
This result shows that the SS module can help increase the diversity of features.

In the following subsections, we further verify the feature extraction capability of the RH-Conv block by comparing it with the other three kinds of feature extraction blocks and then evaluate the multi-grained attention (MGA) module. 

\subsection{Comparison Among Four Kinds of CNN Blocks}
Table~\ref{table2} shows the performance of MGA-Net with four different CNN feature extraction blocks introduced in Section 2.1. 
The “RV-Conv” can achieve better performance compared with “V-Conv”. This may be because introducing residual connection can preserve more of the original features, resulting in a better performance. 
Compared with “RV-Conv”, “RH-Conv” can achieve better performance. It reveals that the combination of CNNs with 1$\times$3 and 3$\times$1 kernels could enhance the feature extraction capability compared with vanilla CNN, especially when serially using asymmetric convolution (1$\times$3, 3$\times$1) and 3$\times$3 convolution. 
Finally, compared with “V-Conv”, the performance on both datasets are increased significantly when the network adopted “RH-Conv”. Especially when focusing on the EB-F1 score, the performance is improved by 1.21$\%$ on the validation and 1.6$\%$ public set.

\begin{table}[] \centering
	\caption{Evaluation of the multi-grained attention module.}
	\renewcommand\arraystretch{1.2}
	\renewcommand\tabcolsep{4.0pt}
	\label{table3}
	\begin{tabular}{l|ll|ll}
		\Xhline{0.8pt}
		\multicolumn{1}{c|}{\multirow{2}{*}{\textbf{Method}}} & \multicolumn{2}{c|}{\textbf{Validation}} & \multicolumn{2}{c}{\textbf{Public}} \\ \cline{2-5} 
		\multicolumn{1}{c|}{}                                 & \textbf{EB-F1}      & \textbf{PSDS}      & \textbf{EB-F1}   & \textbf{PSDS}    \\ \hline
		MGA-Net(Fine-Coarse)                                  & 53.09               & 0.709              & 56.48            & 0.738            \\
		MGA-Net(Coarse-Fine)                                  & \textbf{53.27}      & 0.709              & \textbf{56.96}   & 0.739            \\ \hline \midrule
		\quad -Global                                               & 52.93               & \textbf{0.711}     & 56.78            & \textbf{0.748}   \\
		\quad -Local                                                & 51.91               & 0.705              & 55.59            & 0.734            \\
		\quad -Global-Local                                         & 50.69               & 0.696              & 54.95            & 0.738            \\
		\quad -Frame level                                          & 50.45               & 0.698              & 53.60            & 0.730            \\ \Xhline{0.8pt}
	\end{tabular}
\end{table}

\subsection{Evaluation of Multi-grained Attention Module}
We also investigated the effectiveness of the proposed multi-grained attention module, as shown in Table~\ref{table3}. We firstly explore the feature learning patterns from coarse-level to fine-level (Coarse-Fine) and from fine-level to coarse-level (Fine-Coarse), as shown in Fig.~\ref{fig4}.
The results show that the feature learning pattern from coarse-level to fine-level is slightly better than that from fine-level to coarse-level. Therefore, we adopt the Coarse-Fine feature learning pattern in the following experiments. 

We then investigated how much the proposed global/local or frame-level context modeling contributes to the MGA-Net. As shown in Table~\ref{table3},
when the global context modeling is removed, the performance of SED is only sightly decreased on the EB-F1 metric. 
When the local context modeling is removed, the performances on both datasets are all decreased.
It seems that local context modeling plays a more critical role than global context modeling in time context modeling. 
When both the global and local context modeling is removed, only frame-level context modeling is used to extract the fine temporal information, the performance on both datasets is further decreased.
Results reveal that it is necessary to first conduct the global context modeling before the local context modeling.
In particular, the EB-F1 score is decreased by 2.4$\%$ on the validation and by 2.8$\%$ on the public set.
It also demonstrates that global and local context modeling plays a vital role in capturing event-specific onset and offset information. 
When the frame-level context modeling is removed while preserving the global and local context modeling, we can see that the performance on both datasets is all decreased.

\section{Conclusions}

In this paper, we propose a multi-grained attention network for sound event detection. Four kinds of CNN feature extraction blocks are investigated, and the RH-Conv block has shown it superior to the vanilla CNN block in obtaining features related to the sound events. 
The spatial shift (SS) module provides a data perturbation and shows its effect on increasing features' diversity. 
In addition,
a multi-grained attention (MGA) module is designed to progressively model the time context information from coarse-level to fine-level. Ablation experiments show that a better performance can be achieved when combining the global, local, and frame-level modeling, clearly demonstrating the effectiveness of the proposed method. In the future, we hope to design more effective feature extraction structures to improve sound event detection performance. 

\section{Acknowledgements}

This work is supported by National Natural Science Foundation of
China (NSFC) (U1903213), Tianshan Innovation Team
Plan Project of Xinjiang (202101642)
\\

\bibliographystyle{IEEEtran}

\bibliography{mybib}


\end{document}